\documentclass[fleqn,twoside]{article}
\usepackage{espcrc2}
\usepackage{graphicx}

\newcommand{\mod}{~{\mathrm{mod}}~}
\newcommand{\dual}{\mbox{}^{\ast}}
\newcommand{\dd}{\mbox{\rm d}}

\newcommand{\Z}{{Z \!\!\! Z}}
\newcommand{\beqn}{\begin{eqnarray}}
\newcommand{\eeqn}{\end{eqnarray}}
\newcommand{\eq}[1]{(\ref{#1})}

\title{
\vspace{-18mm}
\thispagestyle{empty}
\rightline{\small ITEP-LAT/2001-1~~~~~}
\vspace{-2mm}
\rightline{\small KANAZAWA-01-11~~~~~}
\vspace{-2mm}
\rightline{\small {\bf 15} October, 2001~~~~~}
Monopoles and hybrids in Abelian projection of lattice QCD
\thanks{Presented by M.I.P. at Lattice 2001, Berlin.}
\author{P.Yu.~Boyko\address{Institute of Theoretical and Experimental
Physics, B.Cheremushkinskaya 25, Moscow, 117259, Russia},
M.N.~Chernodub${}^{\mathrm{a,}}$\address{Institute for Theoretical Physics,
Kanazawa University, Kanazawa 920-1192, Japan},
A.V.~Kovalenko$^{\mathrm{a}}$, S.M.~Morozov$^{\mathrm{a}}$,
M.I.~Polikarpov$^{\mathrm{a}}$}
}

\begin{document}

\begin{abstract}
We study topological defects constructed from diagonal and non-diagonal
gluons in Abelian projection of
zero temperature
lattice QCD. We compare results obtained in
the quenched and non-quenched vacuum field configurations
and show that the density of hybrids is higher than the density of
monopoles. The density of some hybrids is sensitive to the presence of the
virtual fermions.  
\end{abstract}

\maketitle

\section{INTRODUCTION}
There are a lot of numerical facts
indicating
that
Abelian
monopoles in the Maximal
Abelian (MA) projection of lattice gluodynamics play an important role in
the confinement scenario (see, {\em e.g.}, reviews \cite{review}). The
monopoles are constructed from diagonal elements of the gluon field matrix,
these elements
corresponds to
the gauge field in the Abelian
projection~\cite{AbProj}. The off-diagonal gluons play the role of the
electrically charged
matter fields.

Besides the Abelian monopoles one can construct~\cite{hybrid}
other topologically stable defects ("hybrids") using both diagonal and
off-diagonal gluons in $SU(2)$ lattice gauge theory.

In the present publication we generalize the results of Ref.~\cite{hybrid}
to the $SU(3)$ gauge group
and calculate the density of hybrids in quenched and non-quenched lattice
QCD.

\section{DEFINITION OF HYBRIDS}
The monopole current for $SU(2)$ lattice gauge theory is defined as
follows:
\beqn
\dual j = \frac{1}{2\pi}\, \dual\dd (\dd\theta \mod 2\pi)\,,
\label{1}
\eeqn
here $\theta$ is the phase of the diagonal element of the link matrix
$U_{ii} = |U_{ii}|\exp{i\theta_i}$, and we use the notations of lattice
differential forms: $\dd\theta$ is
the standard plaquette angle,
"$\dd$" is a differential operator on the lattice and
the current $\dual j$ is an integer-valued one form
(4-vector) on the dual lattice.

The current defined by eq.~\eq{1}
is
\begin{itemize}
\item[{\em a)}]
{\sl invariant} under the $U(1)$ gauge transformations:
$\theta\rightarrow \theta+d\alpha$, $j\rightarrow j$.
\item[{\em b)}]
{\sl topologically stable}:
a
small (but finite)
va\-ria\-tion of $\theta$ does not change the current $j$.
\item[{\em c)}] {\sl integer--valued}, $\dual j_{x,\mu} \in \Z$.
\item[{\em d)}]
{\sl conserved},
$\delta \dual j = \delta \dual \dd(\dd\theta \mod 2\pi)
= 0$,
due to the identities
$\dd^2 \equiv 0$ and $\dd \equiv \dual \delta \dual$.
\end{itemize}

The generalization of the above construction to $SU(3)$ gauge group was done
in Ref.~\cite{SU3}.
Below we
construct the hybrid $SU(3)$ currents,
satisfying
conditions {\em a)-d)}.

Condition {\em a)} demands the gauge invariance of the hybrid currents
under the $U(1)\times U(1)$ Abelian gauge transformations,
$U_{x,\hat{\mu}}\rightarrow T_x U_{x,\hat\mu} T^\dagger_{x+\hat\mu}$,
where the matrix of the Abelian gauge transformation is defined as follows:
\begin{displaymath}
    T_x = \left( \begin{array}{ccc}
	    e^{i\alpha_x} & 0 & 0  \\
	    0 & e^{i\beta_x}  & 0  \\
	    0 & 0 & e^{-i(\alpha_x+\beta_x)}
    	\end{array} \right).
\end{displaymath}
It is easy to see that under this transformation
the phases of the link matrix
are transformed as shown
in Table~1.
\begin{table}[!htb]
\begin{center}
\caption{The $U(1)\times U(1)$ gauge transformation.}
\begin{tabular}{ll}
\hline
Field & Transformed field\\
\hline
$\theta_1$ & $\theta_1 + \alpha_x - \alpha_{x+\hat \mu}$ \\
$\theta_2$ & $\theta_2 + \beta_x - \beta_{x+\hat \mu}$ \\
$\theta_3$ & $\theta_3 - (\alpha_x - \alpha_{x+\hat \mu}) - (\beta_x -
\beta_{x+\hat \mu})$ \\
$\chi_{12}$ & $\chi_{12} + \alpha_x - \beta_{x+\hat \mu}$ \\
$\chi_{21}$ & $\chi_{21} + \beta_x - \alpha_{x+\hat \mu}$ \\
$\chi_{13}$ & $\chi_{13} + \alpha_x + \alpha_{x+\hat \mu} + \beta_{x+\hat \mu}$ \\
$\chi_{31}$ & $\chi_{31} - \alpha_x - \alpha_{x+\hat \mu} - \beta_{x}$ \\
$\chi_{23}$ & $\chi_{23} + \beta_x + \beta_{x+\hat \mu} + \alpha_{x+\hat \mu}$\\
$\chi_{32}$ & $\chi_{32} - \beta_x - \beta_{x+\hat \mu} - \alpha_{x}$ \\
\hline
\end{tabular}
\end{center}
\end{table}

For simplicity we use the "naive" definition of the
monopole current for $SU(3)$ gauge group,
\beqn
\dual j_i = \frac{1}{2\pi}\dd(\dd\theta_i\mod 2\pi).
\label{jsu3}
\eeqn
It occurs that the results for the monopole current density
obtained with the "naive" definition (\ref{jsu3}) coincide with $10\%$
accuracy with the results obtained using the "accurate" definition given in
Ref.~\cite{SU3} (see also the contribution of T. Streuer
in these Proceedings).
The
definition
of the hybrid current is analogous to
that of
the monopole current. First we construct the $U(1)\times
U(1)$ gauge invariant plaquette variables $p_m$ from $\theta_i$ and
$\chi_{ij}$,
these are the analogues
of $d\theta_i$. Then we define the plaquette angle $P_m$:
\begin{equation}\label{Pm}
	P_m = \sum_k (p_k\mod 2\pi)\,,
\end{equation}
As an example, we
show the construction of $P_m(H1)$ variable explicitly:

\vspace{4mm}
$P_m(H1) =
\vspace{2mm}
$\\
\vspace{4mm}
\includegraphics[scale=0.24]{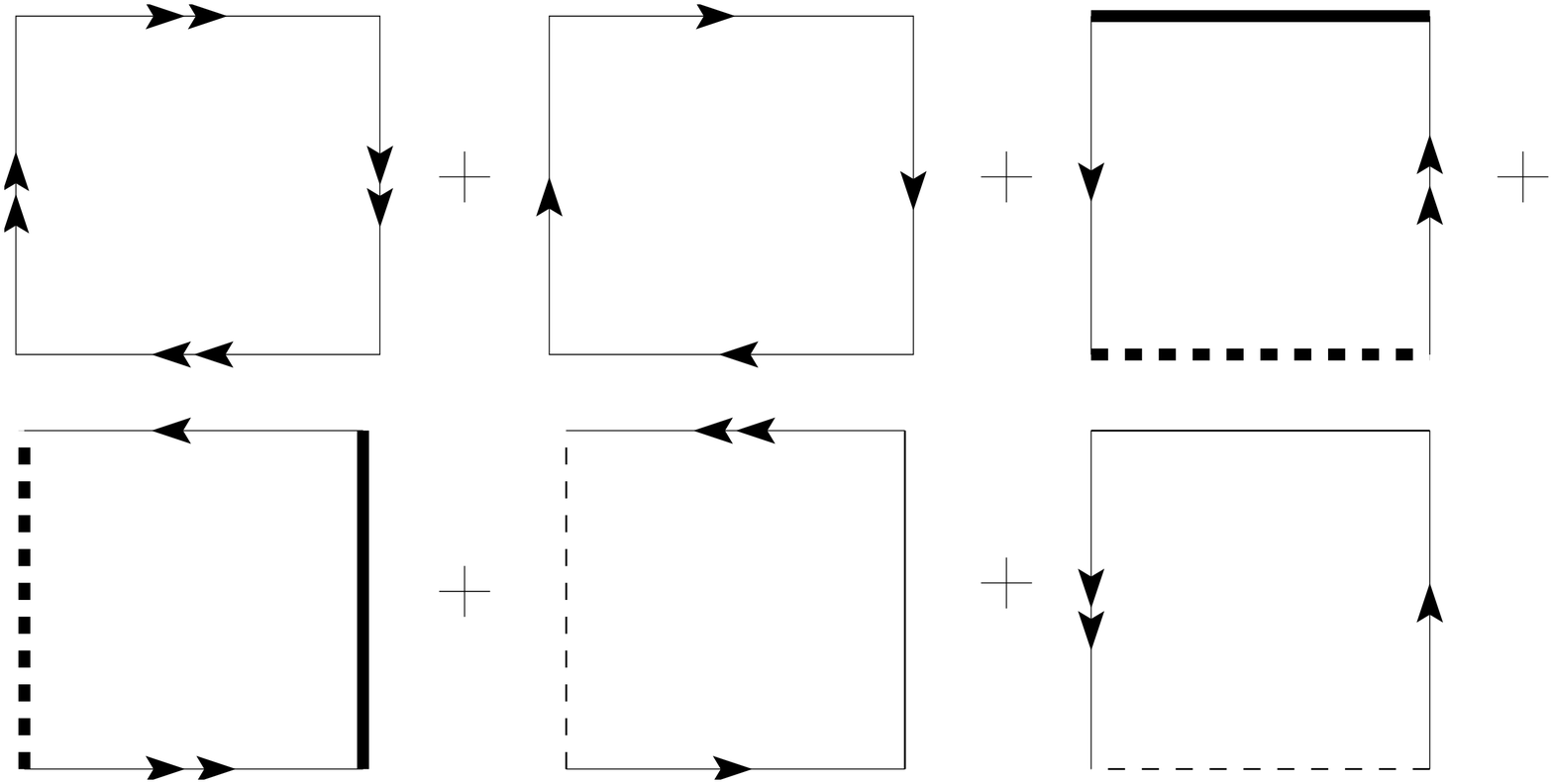}
\vspace{1mm}
\\
Here the variables $\theta$ are shown by solid lines
with arrows; the variables $\chi$
are shown by arrows;
the variables $\chi$ are shown by the solid lines and
the variables $-\chi$ are shown by the dashed lines.
Different species of $\theta$'s differs from each other by the number
of arrows while the species of $\chi$ are distinguished by the width of
the lines.

The 3-forms satisfying condition {\em c)} can be constructed
from the 2--forms $P_m(\theta_i, \chi_{i,j})$  attached to six faces
of the $3D$ cube,
\begin{equation}
\dual {j}_{H1} = \sum_{m=1}^6 P_m(H1)\,.
\end{equation}
We construct $J_{H1}$ in such a way that the hybrids disappear in non-compact
theory (if we disregard $\mod 2\pi$ in eq. (\ref{Pm})). Thus the hybrids are
topologically stable objects
($i.e.$, condition {\em b)} is satisfied). 
In a separate publication we will show that
$\delta j_{H1} = 0$. Thus condition {\em d)} is satisfied for
$j_{H1}$ as well.

There are a lot of types of currents $j_H$ which can be
constructed from $\theta_i$ and $\chi_{ij}$. However, in the
MA projection one can observe a natural hierarchy of the hybrid currents.
Indeed, the MA projection implies the maximization of the modulus of the
diagonal elements, $U_{ii}$, of the link matrix $U$. Therefore
the modulus of the off-diagonal elements $U_{ij}$ is minimized and, as a
result, the diagonal elements are most important for the dynamics.
Since the ordinary monopoles are extracted from diagonal elements, $U_{ii}$,
they are the most important topological defects.

To construct monopoles we use the plaquette angles, $$P_m(\theta_i,
\chi_{ij}) = p_m(\theta_i) \mod 2\pi = p_m^{(0)}.$$ There are 36 $U(1)\times
U(1)$ invariant plaquettes $p_m^{(2)}$ constructed from two variables
$\theta_i$ and two variables $\chi_{ij}$: there exist 24 $U(1)\times U(1)$
invariant plaquettes $p_m^{(3)}$ constructed from one $\theta_i$ and three
variables $\chi_{ij}$ {\em etc}. We expect that hybrids constructed from
$p_m^{(2)}$ are more important for dynamics than those constructed from
$p_m^{(k)}, k = 3,4$. It occurs that there are 3 conserved hybrids $H1, H2,
H3$ constructed from $p_m^{(0)}$ and $p_m^{(2)}$. The construction of $H1$
is outlined above and the construction of $H2$ and $H3$ hybrids is
similar.

\section{NUMERICAL RESULTS}
We have calculated the density of hybrids $H1, H2, H3$ using vacuum
fields~\cite{Streuer} generated in quenched and non-quenched lattice $SU(3)$ QCD
on the lattice $16^3\times32$. The non-quenched configurations are generated for
$O(a)$ improved Wilson fermions, the parameters of simulation are:
$\beta = 5.26, \kappa_{sea} = 0.1345$, the other parameters are given in
Ref.~\cite{lambda}. The quenched configurations are generated for $\beta =
6.0$. This value of $\beta$ is chosen in such a way that the force between
heavy quark-antiquark pair at the distance $r_0 = 0.52 fm$ is the same in
quenched and non-quenched case.

Our numerical results are given in Table~2. The first
raw represents
the density of monopoles~(see talk of T.Streuer on this symposium),
the other lines correspond to the density of hybrids $H1, H2, H3$ discussed
in the previous section.
The last column represents the density of the corresponding hybrids
generated in random ensembles of Abelian fields.

One can draw a few conclusions from Table~2.
The monopole density is much lower than the
corresponding random value, while the hybrids are much more randomized.
The density of $H1$ hybrid is sensitive to the presence of the virtual
quarks. Note, however, that the difference in the densities of hybrids in
quenched and non-quenched vacuum is much smaller than the similar difference
for monopoles.

\begin{table}[!thb]
\caption{The density of the monopoles and hybrids in the quenched and
non--quenched QCD and in random ensembles.}
\begin{center}
\begin{tabular}{llll}
\hline
 & quenched & non-quenched & random \\
\hline
M & 0.0062($\pm$4) & 0.0153($\pm$5) & 0.5023($\pm$4) \\
\hline
H1 & 0.1049($\pm$3) & 0.1136($\pm$3) & 0.2241($\pm$1) \\
H2 & 0.1219($\pm$1) & 0.1245($\pm$2) & 0.2283($\pm$3) \\
H3 & 0.1545($\pm$2) & 0.1551($\pm$3) & 0.3078($\pm$4) \\
\hline
\end{tabular}
\end{center}
\end{table}

\section{NOT RESOLVED QUESTION}
We do not know what magnetic and electric charge carry the hybrids $H1, H2$
and $H3$. It is important to find a linear combination of the hybrid
currents which corresponds to a pure electric current. This current can be
rather sensitive to
the presence of the dynamical fermions.

\section*{ACKNOWLEDGMENTS}
The authors are grateful to F.V.~Gubarev, G.~Schierholz,
T.~Streuer, T.~Suzuki and V.I.~Zakharov for useful discussions.  M.I.P is
partially supported by grants RFBR 00-15-96786, RFBR 01-02-17456, INTAS
00-00111 and CRDF award RP1-2103. M. N. Ch. is supported by JSPS Fellowship
P01023. P.Yu.B., A.V.K. and S.M.M. are very
grateful to Physics Department of Kanazawa University for financial support
on summer 2001.

\end{document}